\documentclass[conference]{IEEEtran}

\usepackage[cmex10]{amsmath}
\usepackage{mathscinet}
\usepackage{amssymb}
\usepackage{fixmath}
\usepackage{algorithm}
\usepackage{algorithmic}
\usepackage[amsmath,thmmarks]{ntheorem}
\usepackage{my_theorems}

\DeclareMathOperator{\Tr}{T}

\DeclareMathOperator{\tr}{tr}
\DeclareMathOperator{\He}{H} 
\DeclareMathOperator{\rank}{rank} 
\DeclareMathOperator{\Expect}{E}

\DeclareMathAlphabet{\mathbit}{OML}{cmr}{bx}{it}

\newcommand{\id}{\mathbf{I}}

\newcommand{\B}[1]{\mathbit{#1}}
\newcommand{\zero}{\mathbf{0}}
\newcommand{\varn}{\sigma_{\eta}^2}

\newcommand{\MAC}{\mathrm{MAC}}
\newcommand{\BC}{\mathrm{BC}}

\newcommand{\SINR}{\mathrm{SINR}}
\newcommand{\lin}{\mathrm{lin}}

\begin{document}
%
\title{A General Rate Duality of the MIMO Multiple Access Channel and the MIMO Broadcast Channel}

\author{\authorblockN{Raphael Hunger and Michael Joham}
\authorblockA{Associate Institute for Signal Processing, Technische
Universit{\"a}t M{\"u}nchen, 80290 Munich, Germany\\
Telephone: +49 89 289-28508, Fax: +49 89 289-28504, Email: \texttt{hunger@tum.de}}}


%


\maketitle

\begin{abstract}
  We present a general rate duality between the multiple access channel
  (MAC)
  and the broadcast channel (BC) which is applicable to systems
  with and without nonlinear interference cancellation.
  Different to the state-of-the-art rate duality 
  \emph{with} interference subtraction from
  Vishwanath et al., the proposed duality is filter-based instead of
  covariance-based and exploits the arising unitary degree of freedom
  to decorrelate every point-to-point link. 
  Therefore, it allows for noncooperative stream-wise decoding
  which reduces complexity and latency. Moreover, the conversion from one
  domain to the other does not exhibit any dependencies during
  its computation making it accessible to a parallel implementation instead
  of a serial one.
  We additionally derive a rate
  duality for systems with multi-antenna terminals 
  when linear filtering \emph{without}
  interference (pre-)subtraction is applied
  and the different streams of a single user are \emph{not} treated
  as self-interference.
  Both dualities are based on a framework already applied
  to a mean-square-error duality between the
  MAC and the BC. Thanks to this novel rate duality, any rate-based optimization
  with linear filtering
  in the BC
  can now be handled in the dual MAC where the arising expressions 
  lead to more efficient algorithmic solutions 
  than in the BC due to the alignment of the channel and 
  precoder indices.
\end{abstract}

%
\IEEEpeerreviewmaketitle

\section{Introduction}

In the past few years, dualities were successfully employed 
as the linking element between the multiple access channel (MAC)
and the broadcast channel (BC). Thanks to various versions of dualities,
many regions of the MAC and the BC were classified to be identical
under a sum-power constraint.
First, the signal-to-interference-and-noise-ratio (SINR)
regions under single-stream transmission per user
were shown to be identical in \cite{schubert02k,Viswanath}.
Second, the mean-square-error (MSE) regions of the MAC and the BC coincide
which has been proven by means of the SINR duality in 
\cite{schubert05c} and later in
\cite{tenenbaum1} or directly in \cite{MeJoHuUt06,HuJoUt08}.
And third, the rate regions of the MAC and the BC under Gaussian signaling
and nonlinear interference cancellation have recently been shown
to be the same, see \cite{Jindal_duality} for the single-antenna case,
\cite{rate_duality} for the multi-antenna case, and
\cite{Weingarten} for the coincidence of the dirty-paper coding rate region
and the capacity region.
A stream-wise duality with power constraints on subsets of antennas
which holds for the optimum filters of a quality-of-service power minimization
was presented in~\cite{YuL07a} for systems with and without nonlinear interference 
cancellation. Due to its stream-wise nature, conversion from one domain to
the dual is complicated since it is not clear how to allocate the SINRs
to the users in case of multi-antenna terminals.
Besides the capability of proving congruency of two regions,
dualities also deliver explicit conversion formulas 
how to switch from one domain to the other. 
In case of the rate duality in \cite{rate_duality}, 
(arbitrary) optimum receive filters generating sufficient
statistics are assumed both in the MAC and in the BC. 
Given transmit covariance matrices in the MAC
are converted to transmit covariance matrices in the dual BC.
Dependencies during these transformations prevent a parallel processing
and force a serial implementation. In addition, the received
data streams have to be decoded jointly which entails a high computational
complexity.

Our contribution in this paper is twofold. First, we 
present a novel rate duality for systems with nonlinear interference
cancellation. One of the key steps involved is the change from
the covariance matrices to the transmit filters by which we gain an isometry as
degree of freedom. This degree of freedom is then used to decorrelate
every point-to-point link thus making a fast parallel stream-wise
decoding possible. As the streams of a single
user now do not interfere with each other, we can employ 
an SINR duality in the style of our MSE duality in \cite{MeJoHuUt06,HuJoUt08}.
Therein, the transmit filters in the dual domain are scaled receivers
of the primal domain and the receive filters 
are scaled
transmitters of the primal domain. We end up with a system of linear equations
to determine these scaling factors.

Our second contribution is a rate duality for linear filtering
applicable to multi-antenna terminals where different streams
of a user are not treated as self-interference.
Up to now, such a duality did not exist and 
hitherto existing dualities for linear filtering 
treat different streams of a user
as virtual users contributing interference
to the user under consideration, see
\cite{schubert02k,Viswanath,Song2007} for example.
In general, the maximum possible rate cannot be obtained when a
duality based on virtual users is applied.
The underlying framework for the proposed linear duality is similar
to the proposed nonlinear duality presented in the following.
Key observation is again the fact that decorrelation allows 
for a stream-wise decoding which also achieves
the rate that is possible under joint decoding.

\section{System Model}

Two systems are considered, namely the MAC where $K$ multi-antenna users
send their data to a common base station which is equipped with $N$
antennas, and the BC where the signal flow is reversed, i.e.,
the base station serves the users. In the former case the transmission
between the $k$th user and the base station is described
by the channel matrix $\B{H}_k\in\mathbb{C}^{N\times r_k}$ 
with $r_k$ denoting the number of transmit antennas at user~$k$.
The BC link, however, is characterized by the Hermitian channel matrix
$\B{H}_k^{\He}$. User $k$ multiplexes $L_k$ data streams.
If interference cancellation is applied in the MAC, we
assume for the sake of readability that the decoding order
is chosen such that user~$1$ is decoded last, whereas the reversed
encoding order is chosen in the BC, i.e., user~$1$ is precoded first.
For different sortings, the users have to be relabeled correspondingly.
Under these assumptions, the rate of user~$k$ in the MAC with
nonlinear interference cancellation reads as~\cite{Yu04}
\begin{equation}
  R_k^{\MAC} = \log_2\frac{\big|\varn\id_N +\sum_{\ell \leq
  k}\B{H}_\ell\B{Q}_\ell\B{H}_{\ell}^{\He}\big|}
  {\big|\varn\id_N +\sum_{\ell <
  k}\B{H}_\ell\B{Q}_\ell\B{H}_{\ell}^{\He}\big|},
  \label{MAC_rate}
\end{equation}
where $\varn$ is the noise variance per antenna and
$\B{Q}_\ell\in\mathbb{C}^{r_{\ell}\times r_{\ell}}$ denotes the transmit
covariance matrix of user~$\ell$. Contrary, user $k$'s
rate in the BC with nonlinear dirty paper coding is~\cite{rate_duality}
\begin{equation}
  R_k^{\BC} = \log_2\frac{\big|\varn\id_{r_k}+\B{H}_k^{\He}\sum_{\ell\geq
  k}\B{S}_\ell\B{H}_k\big|}{\big|\varn\id_{r_k}+\B{H}_k^{\He}\sum_{\ell>
  k}\B{S}_\ell\B{H}_k
  \big|},
  \label{BC_rate}
\end{equation}
where $\B{S}_\ell\in\mathbb{C}^{N\times N}$ is the BC transmit covariance
matrix of user~$\ell$.
If only linear filtering without interference subtraction is applied,
user~$k$ experiences interference from all other users.

\section{Rate Duality for Systems Utilizing Interference Subtraction}


\subsection{Benefits of the Rate Duality with Interference Cancellation}

Besides the ability to show congruency between the two
capacity regions, the decisive reason for utilizing the rate duality is
that all rate expressions are concave functions
of the transmit covariance matrices in the MAC but not in the BC.
Moreover, the optimal sorting of the users can easily be obtained in the MAC.
As a consequence, many rate-based maximizations can be solved 
with efficient algorithms converging to the global optimum in the
MAC and afterwards converted to the BC by means of the duality conversion
formulas.

\subsection{State-of-the-Art Duality}
By means of the MAC-to-BC conversion, we illustrate the state-of-the-art
rate duality from \cite{rate_duality}. 
Both in the MAC and in the BC, all rate expressions depend only on the 
transmit covariance matrices and not on the matrix valued receive filters
since they are implicitly assumed to generate sufficient statistics.
Based on these statistics, the $L_k$ data streams of user $k$ have
to be decoded \emph{jointly}. Given a set of transmit covariance matrices
$\{\B{Q}_k\}$ in the MAC which fulfills a total transmit power constraint
and obtains a rate tuple $R_1^{\MAC},\ldots, R_K^{\MAC}$
under the assumption of optimum receive filters,
the duality in \cite{rate_duality} generates a set of transmit covariance matrices $\{\B{S}_k\}$
for the BC that
fulfills the same total transmit power constraint
and achieves the same rate tuple $R_1^{\BC},\ldots,R_K^{\BC}$.
In the BC, optimum receivers 
yielding sufficient statistics are again required and 
all streams of every individual user have to be decoded jointly as well.

Two key methods utilized 
are the \emph{effective channel} and the \emph{flipped channel} idea.
The former one implies that the capacity of a point-to-point MIMO system
with channel matrix $\B{H}$
subject to an additive Gaussian distortion (noise plus independent interference) 
with covariance matrix $\B{X}$ equals the capacity of a point-to-point system
with \emph{effective channel} matrix $\B{L}^{-1}\B{H}$ subject to
additive Gaussian distortion with identity covariance matrix
if $\B{X} = \B{L}\B{L}^{\He}$. 
Given an arbitrary effective channel of a point-to-point system, 
a system with reversed signal flow and Hermitian effective channel
(\emph{flipped channel}) has the same capacity \cite{Telatar}.
%
%
According to (\ref{MAC_rate}), the rate of user~$k$ in the MAC can
be expressed as
\begin{equation}
  R_k^{\MAC} = \log_2
        \left|\id_N+\B{X}_k^{-1}\B{H}_k\B{Q}_k\B{H}_k^{\He}\right|,
  \label{MAC_rate_with_X}	
\end{equation}
with the substitution 
$\B{X}_k=\varn\id_N+\sum_{\ell=1}^{k-1}\B{H}_\ell\B{Q}_\ell\B{H}_{\ell}^{\He}$.
Introducing the Cholesky
decomposition $\B{X}_k=\B{L}_k\B{L}_k^{\He}$, applying the determinant 
equality $|\id_a+\B{AB}|=|\id_b+\B{BA}|$ for arbitrary 
$\B{A}$ and $\B{B}$ of appropriate dimensions, 
and inserting two identity matrices 
$\id_{r_k}=\B{F}_k^{-1}\B{F}_k=\B{F}_k^{\He}\B{F}_k^{-\He}$, (\ref{MAC_rate_with_X})
can be expressed as 
\[
    R_k^{\MAC} =  \log_2\left|\id_N\!+\!\B{L}_k^{-1}\B{H}_k\B{F}_k^{-1}\B{F}_k\B{Q}_k\B{F}_k^{\He}
        \B{F}_k^{-\He}\B{H}_k^{\He}\B{L}_k^{-\He}\right|.
\]
Now, $\B{L}_k^{-1}\B{H}_k\B{F}_k^{-1}$ can be regarded as the effective
channel for the covariance matrix $\B{F}_k\B{Q}_k\B{F}_k^{\He}$. How
$\B{F}_k$ must be chosen will be clarified below.
Flipping the channel, outcomes in \cite{rate_duality} ensure the existence of
a covariance matrix $\B{Z}_k\in\mathbb{C}^{N\times N}$ with
\begin{equation}
  \begin{split}
     R_k^{\MAC}& =\log_2\left|\id_{r_k}+\B{F}_k^{-\He}\B{H}_k^{\He}\B{L}_k^{-\He}
    \B{Z}_k \B{L}_k^{-1}\B{H}_k\B{F}_k^{-1}\right|,\\
    \tr(\B{Z}_k) & \leq \tr(\B{F}_k\B{Q}_k\B{F}_k^{\He}).
  \end{split}    
  \label{flipped_MAC_rate}
\end{equation}
The rate of user~$k$ in the BC is (cf. Eq.~\ref{BC_rate})
\begin{equation}
  \begin{split}
    R_k^{\BC} &=\log_2\left|\id_{r_k}+\B{Y}_k^{-1}\B{H}_k^{\He}\B{S}_k\B{H}_k\right|\\
    & = \log_2\left|\id_{r_k}+\B{F}_k^{-\He}\B{H}_k^{\He}\B{S}_k\B{H}_k\B{F}_k^{-1}\right|,
  \end{split}    
  \label{BC_rate_with_Y}
\end{equation}
with the substitution 
$\B{Y}_k\! =\! \varn\id_{r_k}\!+\!\sum_{\ell=k+1}^K\B{H}_k^{\He}\B{S}_{\ell}\B{H}_k\!=\!\B{F}_k^{\He}\B{F}_k$.
Equality between $R_k^{\MAC}$ in (\ref{flipped_MAC_rate}) 
and $R_k^\BC$ in (\ref{BC_rate_with_Y})
holds, if
\begin{equation}
  \B{S}_k = \B{L}_k^{-\He}\B{Z}_k\B{L}_k^{-1}.
\end{equation}
Implicitly, $\B{Z}_k$ depends on $\B{F}_k$ as will be shown soon.
Thus, $\B{S}_k$ depends on $\B{Y}_k$ which itself is a function of all
$\B{S}_\ell$ with $\ell>k$. These dependencies require that
$\B{S}_k$ has to be computed before $\B{S}_{k-1}$ and consequently,
one has to start with the computation of $\B{S}_K$ followed by
$\B{S}_{K-1},\ldots,\B{S}_1$.

It remains to determine the matrices $\B{Z}_k \ \forall k$.
Introducing the reduced \emph{singular-value-decomposition} (rSVD)
\begin{equation}
  \B{L}_k^{-1}\B{H}_k\B{F}_k^{-1} = \B{U}_k\B{D}_k\B{V}_k^{\He} \in \mathbb{C}^{N\times r_k}
  \label{rSVD}  
\end{equation}
with the two (sub-)unitary matrices $\B{U}_k\in\mathbb{C}^{N\times\rank(\B{H}_k)}$
and $\B{V}_k\in\mathbb{C}^{r_k\times \rank(\B{H}_k)}$,
the matrix $\B{Z}_k$ reads as
\begin{equation}
  \B{Z}_k =  \B{U}_k\B{V}_k^{\He}\cdot
             \B{F}_k\B{Q}_k\B{F}_k^{\He}\cdot
	     \B{V}_k\B{U}_k^{\He}.
  \label{flipped_matrix}	     
\end{equation}
The proof
for the sum-power conservation
can be found
in~\cite{rate_duality}.
%
%
From the MAC-to-BC conversion, it can be concluded that every rate tuple in the
MAC can also be achieved in the dual BC.
Conversely, the transformation from the BC to the MAC which follows from the
same framework, states that every rate tuple in the BC can also be achieved in the
MAC. Hence, the duality of these two domains is proven and as a consequence, 
their capacity regions are congruent.
Summing up, the state-of-the-art rate duality including interference 
cancellation is serial in two senses: First, it requires a serial implementation of the covariance
matrix conversion due to the dependencies of $\B{S}_k$ on $\B{S}_\ell$ with $\ell>k$.
Second, the application of the duality requires that the different streams associated to a 
user are decoded jointly or, at the best, in a serial fashion.

%


\subsection{Proposed Filter-Based Duality}

The previously described state-of-the-art rate duality
is mainly deduced from information theoretic considerations, where
optimum receivers generate sufficient statistics and capacity is
achieved via joint decoding with inter- and intra-user 
successive interference cancellation.
Approaching from a signal processing point of view
enables us to derive a novel intuitive duality of low complexity.
Switching from arbitrary sufficient
statistics generating optimum receivers to MMSE receivers, we are able
to express all rates in terms of error covariance matrices, which
in turn only depend on the transmit covariance matrices, i.e., on the
outer product of the precoding filters. 
The remaining degree of freedom is a unitary 
rotation and we utilize this isometry in order to decorrelate 
every single point-to-point link. Doing so, the error covariance matrix
becomes diagonal and capacity is achieved with \emph{separate} stream-wise
decoding making intra-user interference cancellation superfluous.
The fact that stream-wise encoding/decoding achieves capacity has already
been observed in \cite{Viswanath,Tse_capacity}. There, however, intra-user
successive decoding must be applied and all streams are decoded one by one.

As all rates can now be expressed as functions of the SINRs
of the individual streams, we apply a low-complexity SINR duality
in the style of our MSE duality in \cite{HuJoUt08,MeJoHuUt06}. In a nutshell, the scaled MMSE receivers
are used as precoders in the dual domain and scaled precoding filters serve
as the receive filters in the dual domain. This dual domain features the
same SINR values as the original one and therefore achieves the same user rates.
In the following, we give an elaborate derivation of the MAC-to-BC conversion.

\subsubsection{Derivation}
Assuming that every MAC covariance matrix
$\B{Q}_k=\B{T}_k\B{T}_k^{\He}$ is
generated by the precoder $\B{T}_k\in\mathbb{C}^{r_k\times L_k}$,
the symbol estimate of user~$k$ in the MAC is
\[
  \hat{\B{s}}_k=\B{G}_k\Big[\B{H}_k\B{T}_k\B{s}_k + 
  \sum_{\ell>k}\B{H}_\ell\B{T}_\ell\B{s}_\ell + 
  \sum_{\ell<k}\B{H}_\ell\B{T}_\ell\B{s}_\ell + \B{\eta}\Big],
\]
where $\B{G}_k$ denotes the receive filter of user~$k$,
$\B{s}_k$ its data vec\-tor with identity covariance matrix, and $\B{\eta}$ the additive noise.
Since interference caused by users 
$l>k$ 
is removed by successive interference cancellation, the MMSE receiver
for user~$k$ is
\begin{equation}
  \B{G}_k=\B{T}_k^{\He}\B{H}_k^{\He}
  \Big(\sum_{\ell\leq k}\B{H}_\ell\B{T}_\ell\B{T}_{\ell}^{\He}\B{H}_{\ell}^{\He}
  +\varn\id_N\Big)^{-1}.
  \label{MMSE_receiver}
\end{equation}
Using (\ref{MMSE_receiver}) and the matrix-inversion lemma, 
the MMSE error covariance matrix 
$\B{C}_k\!=\!\Expect [(\B{s}_k\!-\!\hat{\B{s}}_k)(\B{s}_k\!-\!\hat{\B{s}}_k)^{\He}]$ 
reads as
\begin{equation}
    \B{C}_k  = \id_{L_k}\!-\B{G}_k\B{H}_k\B{T}_k
   = \left[\id_{L_k}\!+\!\B{T}_k^{\He}\B{H}_k^{\He}
   \B{X}_k^{-1}\B{H}_k\B{T}_k\right]^{-1}\!,
  \label{MMSE_cov_matrix}
\end{equation}
with
$\B{X}_k=\varn\id_N+\sum_{\ell=1}^{k-1}\B{H}_\ell\B{T}_\ell\B{T}_{\ell}^{\He}\B{H}_{\ell}^{\He}$.
The rate of user~$k$ can be expressed
in terms of its error covariance matrix
\begin{equation}
  R_k^\MAC = \log_2|\B{C}_k^{-1}|=-\log_2|\B{C}_k|,
\end{equation}
cf.~(\ref{MAC_rate_with_X}).
Note that the rate of user~$k$ is invariant
to a unitary matrix $\B{W}_k$ multiplied from the right hand side to
$\B{T}_k$ yielding $\B{T}_k^{\prime}=\B{T}_k\B{W}_k$. 
Moreover, the rate expressions of other users only depend on the transmit
covariance matrices and not on the filters themselves therefore being also
invariant to this isometry. Last but not least, the transmit power
$\tr(\B{Q}_k)=\tr(\B{T}_k\B{T}_k^{\He})=\tr(\B{T}_k^{\prime}\B{T}_k^{\prime\He})$ is invariant under this 
isometry~$\B{W}_k$. Although $\B{W}_k$ does not influence the interference
covariance matrix experienced by any other user, it can be used as
a spatial decorrelation filter for every point-to-point link which in 
conjunction with the MMSE
receiver $\B{G}_k^\prime = \B{W}_k^{\He}\B{G}_k$ diagonalizes the error-covariance
matrix $\B{C}_k$. To this end, $\B{W}_k$ must be chosen as the
eigenbasis of $\B{G}_k\B{H}_k\B{T}_k$ which is also the eigenbasis of
$\B{T}_k^{\He}\B{H}_k^{\He}\B{X}_k^{-1}\B{H}_k\B{T}_k$. Due to the decorrelation, 
all point-to-point links from
the users to the base station achieve capacity without intra-user successive
interference cancellation thus making separate stream decoding possible.
This way, the rate of user~$k$ can be expressed as the sum of the individual
streams' rates, i.e., $R_k^\MAC =\sum_{i=1}^{L_k} R_{k,i}^{\MAC}$, where
\[
  R_{k,i}^{\MAC} = \log_2(1+\SINR_{k,i}^{\MAC}).
\]
Let $\B{t}_{k,i}^{\prime}$ be the $i$th column of $\B{T}_k^\prime$
and $\B{g}_{k,i}^{\prime\Tr}$ be the $i$th row of $\B{G}_k^{\prime}$, then
the general SINR definition in the MAC 
\begin{equation}
    \SINR_{k,i}^{\MAC} =
    \frac{|\B{g}_{k,i}^{\prime\Tr}\B{H}_k\B{t}_{k,i}^\prime|^2}
     {\B{g}_{k,i}^{\prime\Tr}
       \Big(\B{X}_k
        +\sum_{m\neq i}\B{H}_k\B{t}_{k,m}^\prime\B{t}_{k,m}^{\prime\He}
    \B{H}_k^{\He}
        \!\Big) \B{g}_{k,i}^{\prime *}
      }
  \label{SINR_MAC}
\end{equation}
reduces for the special choice of the decorrelation filter $\B{W}_k$~to
\begin{equation}
    \SINR_{k,i}^{\MAC} =
      \frac{|\B{g}_{k,i}^{\prime\Tr}\B{H}_k\B{t}_{k,i}^\prime|^2}
       {\varn\|\B{g}_{k,i}^{\prime}\|_2^2 + \sum_{\ell<k}\sum_{m=1}^{L_\ell}
       |\B{g}_{k,i}^{\prime\Tr}\B{H}_{\ell}\B{t}_{\ell,m}^{\prime}|^2}
    \label{SINR_MAC_simplified}   
\end{equation}
i.e., the summation over~$m$ in the denominator
of~(\ref{SINR_MAC}) vanishes as $\B{G}_k^{\prime}\B{H}_k\B{T}_k^{\prime}$ is diagonal. Inserting 
$\B{G}_k^{\prime}$ into (\ref{SINR_MAC_simplified}) yields
\begin{equation}
  \SINR_{k,i}^{\MAC} = \B{t}_{k,i}^{\prime\He}\B{H}_k^{\He}\B{X}_k^{-1}\B{H}_k\B{t}_{k,i}^{\prime},
\end{equation}
according to the diagonal entries of
$\B{W}_k^{\He}\B{C}_k^{-1}\B{W}_k$, see (\ref{MMSE_cov_matrix}).

In the dual BC with Hermitian channels, dirty paper coding for 
inter-user interference presubtraction is applied with re\-versed order.
The receivers perform a stream-wise decoding based on the outputs
of the receive filters $\B{B}_k \ \forall k$. Given precoders $\B{P}_1,\ldots,\B{P}_K$,
the SINR of user~$k$'s stream $i$ is
\begin{equation}
  \label{SINR_BC}
 \SINR_{k,i}^{\BC} = \frac{|\B{b}_{k,i}^{\Tr}\B{H}_k^{\He}\B{p}_{k,i}|^2}
                            {\B{b}_{k,i}^{\Tr}\Big(
                            \B{Y}_k+
                            \sum_{m\neq i}\B{H}_k^{\He}\B{p}_{k,m}\B{p}_{k,m}^{\He}\B{H}_k
                            \Big)\B{b}_{k,i}^*},
\end{equation}
and the rate of user~$k$ in the BC with stream-wise decoding reads as
$R_k^{\BC} = \sum_{i=1}^{L_k}\log_2(1+\SINR_{k,i}^{\BC})$.
Besides the de\-corre\-lation, the flipping of transmit and receive \emph{filters}
is the core of our duality: Scaled transmit matrices including the decorrelation in the MAC act as receive filters in the
BC and
scaled receivers in the MAC act as transmit filters in the BC:
\vspace{-1mm}
\begin{equation}
  \B{p}_{k,i}=\alpha_{k,i}\B{g}_{k,i}^{\prime*}\quad \text{and}\quad \B{b}_{k,i}
  = \alpha_{k,i}^{-1}\B{t}_{k,i}^{\prime *}.
  \label{filter_conversion}
\end{equation}
Plugging (\ref{filter_conversion}) into the general BC SINR expression
(\ref{SINR_BC})
we obtain by means of the diagonal structure of $\B{G}_k^{\prime}\B{H}_k\B{T}_k^{\prime}$
\[
  \SINR_{k,i}^{\BC} =\frac{\alpha_{k,i}^2|\B{g}_{k,i}^{\prime\Tr}\B{H}_k\B{t}_{k,i}^\prime|^2}
  {\varn\|\B{t}_{k,i}^\prime\|_2^2+\sum_{\ell>k}\sum_{m=1}^{L_\ell}|\B{g}_{\ell,m}^{\prime\Tr}\B{H}_k
      \B{t}_{k,i}^\prime|^2\alpha_{\ell,m}^2}.
  \label{SINR_BC_simplified}
\]
Equating $\SINR_{k,i}^{\BC}$ with the MAC SINR from (\ref{SINR_MAC_simplified}),
we get
\begin{equation}
\vspace{-1mm}
  \begin{split}
    \alpha_{k,i}^2 & \Big[\varn\|\B{g}_{k,i}^{\prime}\|_2^2 + \sum_{\ell<k}\sum_{m=1}^{L_\ell}
       |\B{g}_{k,i}^{\prime\Tr}\B{H}_{\ell}\B{t}_{\ell,m}^{\prime}|^2\Big] \\
       & \quad -      
      \sum_{\ell>k}\sum_{m=1}^{L_\ell}\alpha_{\ell,m}^2|\B{g}_{\ell,m}^{\prime\Tr}\B{H}_k\B{t}_{k,i}^{\prime}|^2
       =\varn \|\B{t}_{k,i}^\prime\|_2^2,
  \end{split}
  \label{conversion_equality}
\end{equation}
which needs to hold for all users $k$ and all streams $i\in\{1,\ldots,L_k\}$ thus generating the
system of linear equations
\begin{equation}
  \B{M}\!\cdot\!\big[\alpha_{1,1}^2,\ldots,\alpha_{K,L_K}^2\big]^{\Tr} = 
   \varn \big[\|\B{t}_{1,1}^\prime\|^2_2,\ldots,\|\B{t}_{K,L_K}^\prime\|_2^2\big]^{\Tr}
   \label{linear_SOE}
\end{equation}
with the $\sum_{k=1}^K L_k \times \sum_{k=1}^K L_k$ block upper triangular matrix 
\begin{equation}
  \B{M}=\left[\begin{array}{ccc}
    \B{M}_{1,1} & \cdots & \B{M}_{1,K} \\
    \zero  & \ddots  & \vdots \\
    \zero  & \zero & \B{M}_{K,K}
  \end{array}
  \right].
  \label{M_matrix}
\end{equation}
The off-diagonal blocks with $a<b$ read as (cf.~Eq.~\ref{conversion_equality})
\begin{equation}
  \B{M}_{a,b}=-(\B{G}_b^\prime\B{H}_a\B{T}_a^\prime)^{\He}\odot(\B{G}_b^\prime\B{H}_a\B{T}_a^\prime)^{\Tr}
  \in\mathbb{R}^{L_a\times L_b}
  \label{off_diag_block}
\end{equation}
with the \emph{Hadamard} product $\odot$,
and $\B{M}_{a,a}$ is diagonal with
\begin{equation}
  [\B{M}_{a,a}]_{i,i}=\varn\|\B{g}_{a,i}^\prime\|_2^2 -
  \sum_{\ell<a}\sum_{m=1}^{L_\ell}[\B{M}_{\ell,a}]_{m,i}.
  \label{diag_block}
\end{equation}
Since all off-diagonal elements of $\B{M}$ are nonpositive and all diagonal elements are nonnegative,
$\B{M}$ is a \emph{Z-matrix} \cite{nonnegative_matrices}. For $\varn>0$, $\B{M}$ is column diagonally dominant. 
So, $\B{M}$
is an \emph{M-matrix} such that its inverse exists with nonnegative entries~\cite{nonnegative_matrices}
yielding valid solutions $\alpha_{k,i}^2\geq 0$. Because of the block upper triangular structure of $\B{M}$
we can quickly solve for $\alpha_{1,1}^2,\ldots,\alpha_{K,L_K}^2$ via
back-substitution,
in particular since the diagonal blocks $\B{M}_{k,k}$ are diagonal matrices. 
Note that a rank-deficient precoder $\B{T}_m$ manifests in
zero columns and zero rows in $\B{M}$ which have to be removed
before inversion. The respective $\alpha_{m,\cdot}^2$ and 
$\|\B{t}_{m,\cdot}^\prime\|_2^2$ in (\ref{linear_SOE}) also have to be removed,
and finally,
$\B{p}_{m,\cdot}=\zero$ and $\B{b}_{m,\cdot}=\zero$ must be chosen.
\\
Summing up the rows of (\ref{linear_SOE}), we obtain
\begin{equation}
  \sum_{k=1}^K\sum_{i=1}^{L_k}\underbrace{\alpha_{k,i}^2\|\B{g}_{k,i}^\prime\|_2^2}_{
  \|\B{p}_{k,i}\|_2^2}\varn =\varn \sum_{k=1}^K\sum_{i=1}^{L_k}\|\B{t}_{k,i}^\prime\|_2^2,
  \label{power_conservation}  
\end{equation}
stating that the dual BC consumes the same power as the MAC.
Thus, the same or larger (if MMSE receivers are chosen for
$\B{B}_1,\ldots,\B{B}_K$) rates can be achieved in the dual BC as in the primal
MAC under the same transmit power constraint.
The reverse direction of the duality transforming BC filters to the MAC
can be handled with the same framework. Due to its similarity, we skip
its derivation. From this direction of the duality, it follows that the BC
rate region is a subset of the MAC capacity region. In combination
with the former result of the MAC-to-BC conversion stating that the MAC capacity region is a subset of
the BC rate region, the following theorem 
becomes evident with the aid of \cite{Weingarten}
(cf.~\cite{rate_duality}):
\vspace{-1.45mm}
\Theorem{congruency}{section}
{
  The capacity regions of the MAC and the BC are congruent
  under a sum-power constraint.
}
As a consequence, any optimization in the BC can be solved in the MAC,
which offers concave rate expressions suitable for efficient globally convergent
algorithms. Since both capacity regions are congruent, we optimize over the
same region and therefore, do not introduce any suboptimality at this point.
Having found the solution in the MAC we can convert it back to
the BC by means of the duality. Optimality in one domain translates itself
to optimality in the other domain.
The main advantage of the proposed filter-based duality compared to the 
state-of-the-art duality in~\cite{rate_duality} is that both the 
conversion and the decoding in the dual domain can be parallelized
and need not be applied serially as in \cite{rate_duality}.
The computation of the transmit and receive filters features no dependencies
and the decoding process does not require intra-user interference cancellation
or intra-user joint decoding of the streams, all streams of a user can be
decoded independently in parallel.

\subsubsection{Algorithmic Implementation}

Given arbitrary precoding filters $\B{T}_k \ \forall k$ in the MAC,
MMSE 
receivers $\B{G}_k$ are first computed via (\ref{MMSE_receiver}) for all~$k$, see Line~2
in Alg.~\ref{alg:novel_MAC_BC}. The decorrelation filter $\B{W}_k$ is chosen
as the eigenbasis of $\B{G}_k\B{H}_k\B{T}_k$
and afterwards, the transmit and receive filters are adapted,
see Lines~3 and~4.
Thereby, a parallel stream-wise decoding is possible without intra-user
interference cancellation.
Having set up the linear system of equations in (\ref{linear_SOE}) which ensures the 
conservation of the SINRs in the BC, the precoders $\B{P}_k$ and
receivers $\B{B}_k$ are computed with (\ref{filter_conversion}), 
cf.\ Line~8.

\section{Rate Duality for Systems without Interference Subtraction}

In case of linear filtering,
i.e., when nonlinear inter-user interference cancellation is not applied,
user~$k$ experiences interference from 
all other users $\ell\neq k$.
Up to now, a \emph{rate} duality for the linear case without
interference subtraction
does not exist in the literature when multi-antenna terminals are involved
and different streams shall \emph{not} be treated as self-interference.
By jointly decoding the streams in the MAC,
user~$k$ can achieve the rate
\[
     R_k^{\MAC}  = \log_2\Big|\id_N\!+\!
     \big(\sum_{\ell\neq k}\B{H}_\ell\B{Q}_\ell\B{H}_{\ell}^{\He}
      \!+\!\varn\id_N\big)^{-1}\B{H}_k\B{Q}_k\B{H}_k^{\He}\Big|
  \vspace{-1.5mm}
\]
\vspace{-1.6mm}
\begin{equation}
     = -\log_2\big|\id_N - \B{X}^{-1}\B{H}_k\B{Q}_k\B{H}_k^{\He}\big|,\hspace{1.62cm}
    \label{linear_rate}
\end{equation}
with the substitution
$\B{X}\!=\!\varn\id_N\!+\!\sum_{\ell=1}^K\B{H}_\ell\B{Q}_\ell\B{H}_{\ell}^{\He}$.
In con\-trast to systems with interference cancellation
described in the previous section,
this matrix is common to MMSE receivers
\begin{equation}
  \B{G}_k = \B{T}_k^{\He}\B{H}_k^{\He}\B{X}^{-1}
  \label{G_no_SIC}  
\end{equation}
for all users~$k$
and therefore has to be computed only once. 
Applying $\B{G}_k$, user~$k$ experiences the error covariance
matrix
\begin{equation}
  \B{C}_k = \id_{L_k} - \B{T}_k^{\He}\B{H}_k^{\He}\B{X}^{-1}\B{H}_k\B{T}_k,
\end{equation}  
which is again decorrelated by the isometry $\B{W}_k$
since the rate $R_k^{\MAC}=-\log_2|\B{C}_k|$ is again invariant under
this unitary degree of freedom.
Choosing $\B{W}_k$ as the
eigenbasis of $\B{T}_k^{\He}\B{H}_k^{\He}\B{X}^{-1}\B{H}_k\B{T}_k$,
we adapt the receive filter $\B{G}_k^\prime=\B{W}_k^{\He}\B{G}_k$ and
the transmit filter $\B{T}_k^\prime = \B{T}_k\B{W}_k$.
Due to the decorrelation, the error covariance matrix $\B{W}_k^{\He}\B{C}_k\B{W}_k$
is diagonalized and all $L_k$ streams of user~$k$ can be decoded
separately yielding the rate $R_k^{\MAC,\lin} = \sum_{i=1}^{L_k}R_{k,i}^{\MAC,\lin}$,
with the rate
\begin{equation}
  R_{k,i}^{\MAC,\lin} = \log_2(1+\SINR_{k,i}^{\MAC,\lin})
\end{equation}
of user $k$'s stream~$i$.
Its SINR now reads as
\begin{equation*}
    \SINR_{k,i}^{\MAC,\lin} =
      \frac{|\B{g}_{k,i}^{\prime\Tr}\B{H}_k\B{t}_{k,i}^\prime|^2}
       {\varn\|\B{g}_{k,i}^{\prime}\|_2^2 + \sum_{\ell\neq k}\sum_{m=1}^{L_\ell}
       |\B{g}_{k,i}^{\prime\Tr}\B{H}_{\ell}\B{t}_{\ell,m}^{\prime}|^2}.
\end{equation*}
We apply the same rule for finding the precoding and receive filters
$\B{P}_k$ and $\B{B}_k$ of user~$k$ in the BC as we do in case
of inter\-ference cancellation, i.e.,
  $\B{p}_{k,i}=\alpha_{k,i}\B{g}_{k,i}^{\prime*}$ and 
  $\B{b}_{k,i} =\alpha_{k,i}^{-1}\B{t}_{k,i}^{\prime *}$,
see (\ref{filter_conversion}).
With these transformations, the BC SINR reads as
\begin{equation*}
  \SINR_{k,i}^{\BC,\lin} =\frac{\alpha_{k,i}^2|\B{g}_{k,i}^{\prime\Tr}\B{H}_k\B{t}_{k,i}^\prime|^2}
      {\varn\|\B{t}_{k,i}^\prime\|_2^2+\sum_{\ell\neq k}
      \sum_{m=1}^{L_\ell}|\B{g}_{\ell,m}^{\prime\Tr}\B{H}_k
      \B{t}_{k,i}^\prime|^2\alpha_{\ell,m}^2}.
  \label{SINR_BC_simplified_2}
\end{equation*}
Equating the BC and MAC SINRs yields
the system of linear equations (\ref{linear_SOE}),
where the 
matrix $\B{M}$ is not block upper triangular as in (\ref{M_matrix}), since inter-user
interference cancellation is not applied:
\vspace{-3mm}
\begin{equation}
  \B{M}=\left[\begin{array}{ccc}
    \B{M}_{1,1} & \cdots & \B{M}_{1,K} \\
    \vdots &  \ddots  & \vdots \\
    \B{M}_{K,1} & \cdots & \B{M}_{K,K}
  \end{array}
  \right].
  \label{M_matrix_2}
\end{equation}
For this reason, (\ref{linear_SOE}) is solved via \emph{LU-factorization} 
\cite[Section 3.2.5]{Golub} and forward-backward
substitution.
The diagonal blocks of $\B{M}$ are diagonal matrices
with diagonal entries
\vspace{-1mm}
\begin{equation}
  [\B{M}_{a,a}]_{i,i}=\varn\|\B{g}_{a,i}^\prime\|_2^2 -
  \sum_{\ell\neq a}\sum_{m=1}^{L_\ell}[\B{M}_{\ell,a}]_{m,i},
  \label{diag_block_2}
  \vspace*{-1mm}
\end{equation}
such that $\B{M}$ is again an \emph{M-matrix} satisfying the
power conservation equation (\ref{power_conservation}).
With slight modifications, Alg.~\ref{alg:novel_MAC_BC} can be used to
perform the MAC-to-BC conversion without nonlinear inter-user
interference cancellation. In Line~2, $\B{G}_k$ 
must be computed according to (\ref{G_no_SIC}),
and in Line~7, the matrix $\B{M}$ follows from (\ref{M_matrix_2}),
(\ref{off_diag_block}), and
(\ref{diag_block_2}).
Again, the converse direction of the duality underlies the same framework
and completes the proof of the duality in case of linear filtering without 
inter-user interference cancellation:
\vspace{-1mm}
\Theorem{congruence_linear}{section}
{
  The MIMO MAC and the MIMO BC share the same rate region under linear filtering
  and a sum-power constraint both for separate and joint
  de-/encoding of each user's data streams.
}
\vspace{-1mm}
This novel rate duality for systems without interference cancellation allows us to 
convert any rate-based optimization from the BC to the MAC without loss
of optimality. An immediate benefit is that we can switch from the
rate expression
\begin{equation*}
    R_k^{\MAC,\mathrm{interference}} = -\log_2 \prod_{i}
   \big[\id_{L_k}-\B{T}_k^{\He}\B{H}_k^{\He}\B{X}^{-1}\B{H}_k\B{T}_k\big]_{i,i}
\end{equation*}
with separate stream decoding and hence self-interference
to the one in (\ref{linear_rate}) with joint stream decoding
\begin{equation*}
  R_k^{\MAC,\lin} = -\log_2 \big|\id_{L_k}-\B{T}_k^{\He}\B{H}_k^{\He}\B{X}^{-1}\B{H}_k\B{T}_k\big|,
\end{equation*}
which is always larger than or equal to $R_k^{\MAC,\mathrm{interference}}$.
Moreover, the channel and precoder indices are aligned in the MAC,
see (\ref{linear_rate}), whereas they aren't in the BC. Although
(weighted) sum-rate maximization remains a nonconcave maximization in the
MAC, the aforementioned indices alignment allows for simpler expressions
and reduced-complexity algorithms. 
Last but not least, MAC precoders 
are characterized by only $\sum_{k=1}^K r_k^2$ variables instead of 
$N\sum_{k=1}^K r_k$ in the BC.
Summing up, solving rate based optimizations with linear filtering in the MAC
and applying the proposed duality is more efficient than solving the problem
in the BC.
\vspace{-0.5mm}
\begin{algorithm}[!t]
  \begin{algorithmic}[1]
    \FOR{$k=1:K$}
             \STATE $\B{G}_k\leftarrow \B{T}_k^{\He}\B{H}_k^{\He}
	       \big(\sum_{\ell\leq k}\B{H}_\ell\B{T}_\ell\B{T}_{\ell}^{\He}\B{H}_{\ell}^{\He}
	       +\varn\id_N\big)^{-1}$ 
      \STATE $\B{W}_k\leftarrow \operatorname{eigenbasis}(\B{G}_k\B{H}_k\B{T}_k)$ \hfill \emph{decorrelation matrix}
      \STATE $\B{G}_k^\prime \leftarrow \B{W}_k^{\He}\B{G}_k$ and 
             $\B{T}_k^\prime \leftarrow \B{T}_k\B{W}_k$ \hfill \emph{decorrelate}
    \ENDFOR             
    \STATE set up $\B{M}$ with (\ref{M_matrix}) -- (\ref{diag_block}),
           remove zero columns/rows
    \STATE solve for $\alpha_{1,1}^2,\ldots,\alpha_{K,L_K}^2$ via (\ref{linear_SOE})
    \STATE $\B{p}_{k,i}=\alpha_{k,i}\B{g}_{k,i}^{\prime *}$ \ \ and \ \  
           $\B{b}_{k,i}=\frac{1}{\alpha_{k,i}}\B{t}_{k,i}^{\prime *} \ \ \forall k,\ \  \forall i$
  \end{algorithmic}
  \caption{Novel stream-wise MAC-to-BC conversion.}
  \label{alg:novel_MAC_BC}	
\end{algorithm}

%
%

%






%

\bibliographystyle{IEEEbib}
\bibliography{globecom}

\end{document}